\documentclass[11pt,a4paper]{article}
\usepackage{jheppub}
\usepackage{natbib}
\usepackage{graphicx}
\usepackage{cases}
\usepackage[dvipsnames]{xcolor}
\usepackage{mathtools}
\usepackage{hyperref}
\usepackage{float}
\usepackage{comment}
\usepackage{multirow}
\usepackage[overload]{empheq}

\def\apjl{ApJL}

\definecolor{Blu}{rgb}{0.,0.,1.}

\author[a,b]{Arpan Kar,}
\author[a,b]{Hyomin Kim,}
\author[c]{Sang Pyo Kim,}
\author[a,b]{Stefano Scopel,}
\affiliation[a]{Center for Quantum Spacetime, Sogang University, Seoul 121-742, South Korea}
\affiliation[b]{Department of Physics, Sogang University, Seoul 121-742, South Korea}
\affiliation[c]{Department of Physics, Kunsan National University, Kunsan 54150, Korea}
\emailAdd{arpankarphys@gmail.com}
\emailAdd{hyomin1996@naver.com}
\emailAdd{sangkim@kunsan.ac.kr}
\emailAdd{scopel@sogang.ac.kr}

\title{WIMP constraints from black hole low--mass X--ray binaries}

\abstract{The abnormally fast orbital decay observed in the black hole (BH) Low--Mass X–ray binaries (BH–LMXB) A0620-00 and XTE J1118+480 can be explained by the dynamical friction between Dark Matter (DM) and the companion star orbiting around the low--mass BH ($\simeq$ a few $M_\odot$) of the system. In this case the value of the index $\gamma_{\rm sp}$ of the DM spike surrounding the BH can be pinned down with an accuracy of $\simeq$ a few percent, way better than that for much bigger systems such as the super massive BHs (SMBHs) in the Galactic Center or in M87. We have used data from XTE J1118+480 to put bounds on the WIMP annihilation cross section times velocity $\langle \sigma v\rangle$, assuming that DM annihilation is driven by the $\chi\chi\rightarrow b\bar{b}$ annihilation channel and that it proceeds in $s$--wave. The bounds are driven by the radio synchrotron signal produced by $e^\pm$ final states propagating in the magnetic field in the vicinity of the BH. We find that for DM masses $m_\chi$ up to the TeV scale XTE J1118+480 allows to constrain $\langle \sigma v\rangle$ well below the standard value $\langle\sigma v\rangle_{\rm thermal}$, corresponding to the observed DM relic density in the Universe for a thermal WIMP. On the other hand, for $m_\chi \gtrsim$ 15 GeV, the bounds from the SMBHs in the GC or in M87 do not reach $\langle\sigma v\rangle_{\rm thermal}$ when the very large uncertainties on the corresponding spike indices are taken into account, in spite of potentially producing much larger DM densities compared to XTE J1118+480. Our bounds for XTE J1118+480 have a mild sensitivity on the effect of spatial diffusion (which implies at most a weakening of the bounds of a factor $\lesssim$ 6 at large $m_\chi$). However, diffusion is instrumental in enhancing the sensitivity of the results upon the intensity of the magnetic field. In particular, our bounds rest on the assumption that the magnetic field $B$ reaches the equipartition value $B^{\rm eq}$. We find that a reduction factor of the magnetic field $B^{\rm eq}/B$ larger than about 14 at low $m_\chi$, becoming progressively smaller at higher WIMP masses, would be sufficient to relax the XTE J1118+480 bound to the level of other existing bounds. Recent estimates, albeit not conclusive, may suggest values of $B^{\rm eq}/B$ in BH–LMXB systems as large as 20. This implies that the intensity of the magnetic field in BH–LMXB systems represents the major uncertainty in using them as an alternative to heavier BHs to search for WIMPs.} 

\begin{document}
\hspace*{87.0mm}{CQUeST-2023-0725}\\
\maketitle

\section{Introduction}
\label{sec:introduction}
The thermal decoupling of a Weakly Interacting Massive Particle (WIMP) remains so far the most popular and simple scenario to explain the Cold Dark Matter (CDM) that is believed to account for $\simeq$ 23\% of the total energy density of the Universe \cite{Planck:2018vyg} and is needed to explain structure formation. WIMP phenomenology is characterized by the fact that the same interactions that are assumed to keep WIMPs in thermal equilibrium with Standard Model (SM) particles in the Early Universe and that allow to predict its relic density in agreement to observation through the thermal decoupling mechanism provide an opportunity to detect them experimentally today. In particular, the indirect detection of the WIMP annihilation products 
($\gamma$'s, $\nu$'s, $e^+$, $\bar{p}$) in the halo of our Galaxy is being extensively pursued using both 
ground--based and satellite telescopes \cite{HESS_GC:2022ygk, IceCube:2017rdn, Abazajian:2020tww, Leane:2018kjk, Calore:2022stf}. For a thermal WIMP the reference value for the average of the WIMP annihilation cross section to SM particles time velocity $\langle \sigma v\rangle$ which corresponds to the observed DM density is of the order of $\langle \sigma v\rangle_{\rm thermal}\simeq$ 2.2$\times$10$^{-26}$ cm$^3$ s$^{-1}$~\cite{Thermal_relic_Steigman:2012nb} and is the same driving today the annihilation of WIMP particles in the halo of our galaxy if DM annihilation proceeds in $s$--wave, i.e. if 
$\langle \sigma v \rangle$ is not suppressed by the non--relativistic WIMP velocity $v$.

WIMP indirect detection signals are proportional to the square of the DM local density $\rho_{\rm DM}$, so that they are enhanced wherever $\rho_{\rm DM}$ is large, such as in the Galactic Center (GC), in dense dwarf galaxies hosted by the Milky Way or in astrophysical object like the Sun or the Earth that can accumulate a dense population of DM particles in their interior through gravitational capture. 

As far as the GC is concerned, dissipationless DM-only simulations of halos with masses ranging from dwarf galaxies to rich clusters suggest that at small radii $r$ the DM density follows a near-universal cusped profile diverging as $\simeq r^{-\gamma_c}$ with $\gamma_c\simeq$ 1--1.5~\cite{nfw, moore}, enhancing the expected flux from DM annihilations. In addition, the Supermassive Black Hole (SMBH) harbored at the center of our Galaxy with mass $M_{\rm BH}\simeq 4\times10^6 M_\odot$ can produce a DM spike with an even larger density behaving as $\simeq r^{-\gamma_{\rm sp}}$ with $\gamma_{\rm sp}\simeq$ as large as $\simeq$ 2.5 in the case of adiabatic contraction~\cite{Gondolo_Silk_adiabatic} when $r\rightarrow$ 0. As a consequence, the DM spike produced by the SMBH at the center of the Milky Way has been extensively studied in the literature as one of the most promising ways to detect WIMPs.

The existence of such a DM spike in the center of our Galaxy is however object of debate. For instance, the growth of the BH could have been non adiabatic (as expected if the BH was produced by a merger~\cite{ullio_zhao_2001}). Moreover, the DM halo itself could have experienced a merger~\cite{Merritt_2002}, or the BH could have grown not exactly at the center of the DM halo~\cite{ullio_zhao_2001}, leading to a much shallower inner profile. Also, even if a DM spike is initially formed it can be subsequently smoothed down by DM scattering off  stars~\cite{merritt_2004}. As a consequence, although when taken at face value some of the constraints on the WIMP annihilation cross section obtained from the DM spike in the GC can be very strong~\cite{Fields_gamma_excess_2014, Shapiro_2016}, they cannot be considered robust. Indeed, in~\cite{Fields_gamma_excess_2014, Shapiro_2016} the exponent $\gamma_{\rm sp}$ is treated as a free parameter and the ensuing WIMP indirect signals span several orders of magnitude being very sensitive to it (see, for instance, the bounds reported in Fig.~\ref{fig:sv_mx}).

In the present paper we argue that, in spite of producing much smaller DM densities compared to those potentially within reach of SMBHs, an alternative and complementary strategy to look for an enhanced WIMP annihilation signal is to use the much lighter black holes with $M_{\rm BH}\simeq$ a few $M_\odot$ observed in black hole low--mass X--ray binaries (BH--LMXB's). 

In such systems a companion star with mass $m$ orbits around a BH with mass $M_{\rm BH}$. In particular, the study of the two binaries closest to Earth, A0620-00 and XTE J1118+480, allows to measure with good accuracy the nearly circular orbit of the companion star providing very precise measurements of the physical parameters of the system, including $M_{\rm BH}$, $m$ and the orbital period $P$~\cite{DM_spike_LMBH_2023}. In both cases this leads to orbital decays $\dot{P}$ that are two orders of magnitude faster than what expected from the emission of Gravitational--Wave radiation, $\dot{P}\simeq$ -0.02 ms yr$^{-1}$: $\dot{P}$ = -0.60 $\pm$ 0.08 ms yr$^{-1}$ for A0620-00 and  $\dot{P}\simeq$ -1.90 $\pm$ 0.57 ms yr$^{-1}$ for XTE J1118+480. Interestingly, albeit alternative scenarios have been put forward~\cite{justham_2006, Chen_Lee_2015}, an intriguing explanation of such discrepancy is that the companion star orbit around the BH is slowed down by the dynamical friction with a DM spike~\cite{DM_spike_LMBH_2023}. In this case, at variance with what happens with SMBHs in the center of Galaxies for which $\gamma_{\rm sp}$ is largely uncertain, this allows to pin down the value of $\gamma_{\rm sp}$ in the BH–LMXBs by getting a prediction of $\dot{P}$ in terms of $\gamma_{\rm sp}$ and comparing it to the observational value. Indeed, in the following we will show that the bounds from BH–LMXBs that can be obtained in this way are potentially affected by smaller uncertainties and are competitive or even stronger than those from SMBHs. However, we also argue that in the case of BH–LMXBs the major uncertainty in the bounds comes from the intensity of the magnetic field. 

The plan of the paper is the following. In Section~\ref{sec:signal} we summarize how the expected WIMP indirect signals from BH–LMXBs are calculated: in particular we summarize the procedure used in~\cite{DM_spike_LMBH_2023} to model the dynamical friction of the companion star within the DM spike in Section~\ref{sec:spike_profile}, and the magnetic field (corresponding to the equipartition assumption) in Section~\ref{sec:magnetic_field}, while in Section~\ref{sec:synchrotron_and_IC} we outline how the two main sources of radiation in the radio and X--ray bands (synchrotron radiation and Inverse Compton (IC) signals produced by $e^\pm$'s from DM annihilation) are estimated. We present our quantitative results in Section~\ref{sec:results}, where we provide the bounds on the WIMP annihilation cross section from BH–LMXBs and compare them to those obtained by other techniques. In this Section we make the standard assumption that the magnetic field is equal to the equipartition value. In Section~\ref{app:b_sens} we discuss the sensitivity of our results on the modeling of the magnetic field, and show the implications of values below equipartition, as suggested by some observations of binary systems. Our Conclusions are provided in Section~\ref{sec:conclusions}.

\section{DM annihilation signals from black hole low-mass X-ray binaries}
\label{sec:signal}

\subsection{DM profile around BH–LMXBs}
\label{sec:spike_profile}

Following the dark matter density spike theory, dark matter
redistributes around the black
hole in the BH–LMXB to form a density spike  within the spike radius $r_{\rm sp}$. Following \cite{DM_spike_LMBH_2023}, the DM density profile around a BH–LMXB is modeled as (taking the effect of DM annihilation into account \cite{radio_signal_M87_2015, lacroix_2018, Chan_new_method_annihilation_2023}):

\begin{align}[left ={\rho_{\rm DM}(r) = \empheqlbrace}]
\label{eq:rho_profile}
& 0 \,\,\,\,\,\,\,\,\,\, {\rm for} \,\,\,\, r \leq 2 R_s \, ,\nonumber\\
& \frac{\rho_{\rm sp}(r) \, \rho_{\rm sat}}{\rho_{\rm sp}(r) \, + \, \rho_{\rm sat}} 
\,\,\,\,\,\,\,\,\,\, {\rm for} \,\,\,\, 2 R_s < r \leq r_{\rm sp} \, , \\
& \rho_0 \,\,\,\,\,\,\,\,\,\, {\rm for} \,\,\,\, r > r_{\rm sp} \, , \nonumber
\end{align}

\noindent with 
\begin{equation}
\rho_{\rm sp}(r) = \rho_0 \left( \frac{r}{r_{\rm sp}} \right)^{-\gamma_{\rm sp}} \, ,
\label{eq:rho_spike}
\end{equation}
and
\begin{equation}
\rho_{\rm sat} = \frac{m_\chi}{\langle \sigma v \rangle \, t_{\rm BH}} \, . 
\label{eq:rho_sat}
\end{equation}

\noindent Here $R_s = 2 G M_{\rm BH} / c^2$ is the horizon radius of the BH ($G$ is the gravitational constant), 
whose value for the BH in XTE J1118+480 (with $M_{\rm BH} \simeq 7.46$ $M_\odot$) is 
$R_s \simeq 7\times10^{-13}$ pc. The quantity $\rho_0$ is the local DM density 
whose estimated value at the location of the considered X-ray binary XTE J1118+480 is 
$\rho_0 = 0.34 \pm 0.03$ ${\rm GeV \, cm^{-3}}$ \cite{DM_spike_LMBH_2023}. 
The parameters $\gamma_{\rm sp}$ and $r_{\rm sp}$ are the 
spike index and the spike radius, respectively. We follow the standard assumption, 
$r_{\rm sp} = 0.2 \, r_{\rm in}$, where the radius of inﬂuence $r_{\rm in}$ is obtained from 
(see \cite{DM_spike_LMBH_2023}): 
\begin{equation}
M_{\rm DM} (r \leq r_{\rm in}) = \int^{r_{\rm in}}_0 dr \, 4 \pi r^2 \, \rho_{\rm DM} = 
2 \, M_{\rm BH} \, .
\label{eq:r_in}
\end{equation}

\noindent Using the expression above for XTE J1118+480 we find $r_{\rm in}\simeq$ 7.3 pc  (the effect of annihilation on the total mass $M_{\rm DM}$ is negligible). 
In Eq.~(\ref{eq:rho_sat}) $m_\chi$ and $\langle \sigma v \rangle$ are the mass and 
the annihilation cross section times velocity of the DM particles. 
The age of the Black Hole $t_{\rm BH}$ is not directly known, although one can assume the upper bound 
$t_{\rm BH}\le P/\dot{P}$~\cite{DM_spike_LMBH_2023}.

\begin{figure*}[ht!]
\centering
\includegraphics[width=7.49cm,height=6cm]{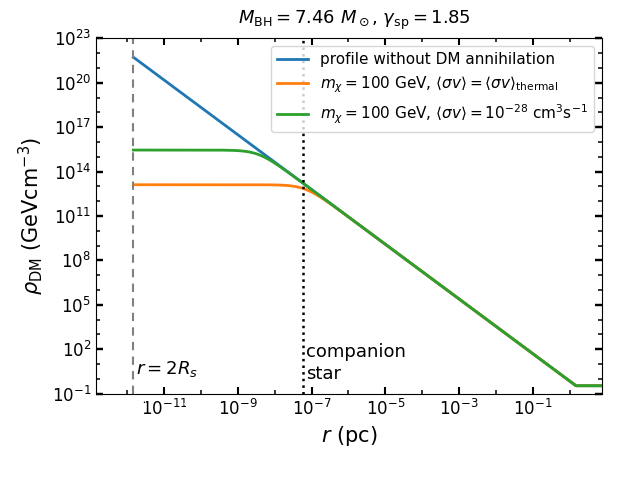}
\caption{DM density distribution as a function of the radial distance from the center of the BH in XTE J1118+480. The blue line shows the density profile without the effect of the DM annihilation (i.e., Eq.~\eqref{eq:rho_profile} when $\langle\sigma v\rangle \to 0$). The orange and green lines show the density profiles if the DM particles have a mass $m_\chi = 100$ GeV and have an annihilation cross section $\langle\sigma v\rangle = {\langle\sigma v\rangle}_{\rm thermal} = 2.2\times10^{-26} \, {\rm cm^3 s^{-1}}$ or 
$\langle\sigma v\rangle = 10^{-28} \, {\rm cm^3 s^{-1}}$, respectively. 
Here the spike index is taken to be $\gamma_{\rm sp} = 1.85$. The black dotted line indicates the location of the companion star (at $r = r_* \simeq 0.0118 \, {\rm AU} \simeq 5.7\times10^{-8} \, {\rm pc}$ \cite{DM_spike_LMBH_2023}).}
\label{fig:rho_r}
\end{figure*}

The spike index $\gamma_{\rm sp}$ is critical for the calculation of expected signals, and, in general, subject to large uncertainties. However in an X-ray binary the accurate measurement of the orbital parameters of the companion star can shed light on it through the dynamical  friction effect~\cite{dynamical_friction_1943}, which consists in the collective gravitational force exerted on a star that is orbiting inside a massive background. In particular, if the companion star in the binary moves inside a DM spike it is followed by a concentration of dark matter particles that slows its orbit due to its gravitational pull. The ensuing orbital decay rate can be directly measured and compared to the corresponding theoretical expectation, which, besides the observed parameters of the 
BH–LMXB~\cite{binary_obs_2008,binary_obs_2011,binary_obs_2013,binary_obs_2014,binary_obs_2017}, and using Eq.~(\ref{eq:rho_profile}) for $\rho_{\rm DM}$, depends on the spike index $\gamma_{\rm sp}$ as the only free parameter~\cite{DM_spike_LMBH_2023}: 
\begin{equation}
\dot{P}\simeq\frac{6\pi q G P \ln q}{(1+q)^2(K/\sin i)}
\left[\frac{G M_{\rm BH}(1+q)P^2}{4\pi^2} \right]^{1/3} \rho_{\rm DM},
    \label{eq:orbital_decay_rate}
\end{equation}

\noindent with $q=m/M_{\rm BH}$ ($M_{\rm BH}$ and $m$ are the masses of the BH and of the companion star, respectively), and $K$ and $\sin i$ the observed radial velocity and orbital inclination. In this way Ref.~\cite{DM_spike_LMBH_2023} obtained for the spike index $\gamma_{\rm sp}$ = 1.71$_{-0.02}^{+0.01}$ for A0620-00 and $\gamma_{\rm sp}$ = 1.85$_{-0.04}^{+0.04}$ for XTE J1118+480. In the following we will use the signal expected from DM annihilation in the X--ray binary J1118+480, which among the two has the largest spike index, to constrain the annihilation cross section times velocity $\langle\sigma v\rangle$, using for $\gamma_{\rm sp}$ the central value $\gamma_{\rm sp}$ = 1.85. Throughout the paper we will assume that DM annihilation proceeds in $s$--wave. Notice that due to the pair-annihilations of the DM particles, in Eq.~(\ref{eq:rho_profile}) the density profile reaches a saturation value $\rho_{\rm sat}$ and creates an annihilation plateau when $r\ll r_{\rm pl}$, with $r_{\rm pl}$ an inner radius where $\rho_{\rm sp}(r_{\rm pl}) = \rho_{\rm sat}$ (see Fig.~\ref{fig:rho_r}). On the other hand the determinations of $\gamma_{\rm sp}$ in Ref.~\cite{DM_spike_LMBH_2023} assume that the companion star is orbiting inside the DM spike, i.e. that the orbital radius of the companion star is larger than $r_{\rm pl}$. This condition depends on $\langle\sigma v\rangle$, so, for consistency, it must be verified {\it a posteriori} once the bound on $\langle\sigma v\rangle$ is determined. Indeed, for $t_{\rm BH}\le P/\dot{P}$ the bounds that will be discussed in Section~\ref{sec:results} will always be strong enough to imply an $r_{\rm pl}$ smaller than the observed orbital radius of the companion star of XTE J1118+480, and so consistent with the determination of $\gamma_{\rm sp}$ from Ref.~\cite{DM_spike_LMBH_2023}.

\subsection{Magnetic field profile around the black hole}
\label{sec:magnetic_field}

Following Ref.~\cite{DM_spike_LMBH_2023} we assume that the BH in the binary and the magnetic field $B$ are spherically symmetric. In particular 
(see \cite{Accreting_BH_Sgr_A*_1992, Bertone_radiation_GC_2002, Regis_Ullio_multiwave_2007, radio_signal_M87_2015, Cermeno:2022rni}) 
we parameterize $B$ near the BH as:
\begin{numcases}{B (r) = }
& $B^{\rm eq} (r)$ \,\,\,\,\,\,\,\,\,\, for \,\,\,\, $r_{\rm H} \leq r < r_{\rm acc}$ \nonumber\\
& $B^{\rm eq} (r_{\rm acc}) \left( \frac{r}{r_{\rm acc}} \right)^{-2}$ 
\,\,\,\,\,\,\,\,\,\, for \,\,\,\, $r \geq r_{\rm acc}$,
\label{eq:B_profile}
\end{numcases}
with 
\begin{equation}
B^{\rm eq} (r) = B^{\rm eq}_{\rm H} \left( \frac{r}{r_{\rm H}} \right)^{-5/4},
\label{eq:B_eq_profile}
\end{equation}
\noindent the magnetic field corresponding to the equipartition assumption (in Section~\ref{app:b_sens} we will discuss smaller values of the magnetic field).
Here, $B^{\rm eq}_{\rm H}$ is the 
magnetic field at the event horizon $r_{\rm H} = R_s = 2 G M_{\rm BH} / c^2$, that can be expressed in terms of the dimensionless accretion rate $\dot{m}$ ($\equiv \dot{M}/\dot{M}_{\rm Eddington}$) as \cite{Accreting_BH_Sgr_A*_1992, gamma_obsv_FermiLAT_2017}
\begin{equation}
B^{\rm eq}_{\rm H} \simeq 4\times 10^{14} \, \dot{m}^{1/2} \, \left(\frac{M_{\rm BH}}{10 M_\odot}\right)^{-1/2} \,\, 
[\mu G] \, .
\label{eq:B_eq_H}
\end{equation}
In Eq.~\eqref{eq:B_profile}, $r_{\rm acc}$ denotes the accretion radius of the BH. For the BH-binary 
XTE J1118+480 we adopt from \cite{binary_obs_2013, MMT_observation_XTE} the value $r_{\rm acc} \simeq 0.6 \, r_*$, with $r_*$ the orbital radius of the companion star of the binary 
($r_* \simeq 0.0118 \, {\rm AU} \simeq 5.7\times10^{-8} \, {\rm pc}$ \cite{DM_spike_LMBH_2023}).
%close to the orbital radius of the companion star of the binary. 
According to \cite{Tetarenko:2015vrn, gamma_obsv_FermiLAT_2017}, the estimated 
value of $\dot{m}$ for XTE J1118+480 is $\dot{m} \simeq 9.92\times10^{-5}$, 
which (with $M_{\rm BH} \simeq 7.46$ $M_\odot$) implies 
$B^{\rm eq}_{\rm H} = (B^{\rm eq}_{\rm H})_0 \simeq 4.5\times 10^{12}$ $\mu G$.

Apart from the magnetic field profile described in Eq.~\eqref{eq:B_profile} (which is centered near the BH), we assume a constant background magnetic field $B_0 = 1$ $\mu G$ which is similar to the value of the large scale Galactic magnetic field at the location of the BH-binary XTE J1118+480 (longitude $\simeq 157.7^{\circ}$, 
latitude $\simeq 62.3^{\circ}$, distance from Earth $\simeq 1.7$ kpc) \cite{Ibarra:2009, Buch:2015}.

\subsection{Synchrotron and Inverse Compton signals} 
\label{sec:synchrotron_and_IC}
Assuming a self-conjugate WIMP the amount of $e^+/e^-$ with energy $E$ produced by 
WIMP annihilations per unit time, volume and energy is given by \cite{multifrequency_coma_cluster_2006, Regis_Ullio_multiwave_2007}: 
\begin{equation}
Q_{e}(r,E) = {\langle \sigma v \rangle} \frac{\rho^2_{\rm DM}(r)}{2 m^2_{\chi}} 
\sum_{F} B_F \frac{dN^F_{e}}{dE} (E,m_{\chi})  \, ,
\label{eq:DM_source_fn}
\end{equation} 
where ${\rho^2_{\rm DM}(r)}/{2 m^2_{\chi}}$ is the number density of annihilating DM pairs 
at the location $r$, $B_F$ is the branching fraction to the primary annihilation channel $F$, while ${dN^F_{e}}/{dE}$ is the $e^+/e^-$ energy spectrum per $F$ annihilation. 
The WIMP primary annihilation channels can be (depending on $m_{\chi}$) 
$e^+e^-$, $\mu^+\mu^-$, $\tau^+\tau^-$, $b\bar{b}$,  $t\bar{t}$, $\gamma\gamma$, $W^+W^-$, $ZZ$, etc. In the literature it is customary to present the bounds on $\langle \sigma v\rangle$ by assuming either the $\chi\chi\rightarrow b\bar{b}$ or the $\chi\chi\rightarrow W^+W^-$ annihilation channel. For definiteness, and with the goal of comparing in Section~\ref{sec:results} the constraints  from  XTE J1118+480 to those from other systems, in the following  we will assume that WIMP annihilation is driven by $\chi\chi\rightarrow b\bar{b}$. The energy spectra ${dN^F_{e}}/{dE}$ for the electrons and positrons 
produced in this annihilation channel are obtained from the files generated using $\mathtt{Pythia}$~\cite{Sjostrand:2006za} by $\mathtt{PPPC4DMID}$~\cite{Cirelli:2010xx,Ciafaloni:2010ti,Cirelli}.

After production the $e^\pm$'s propagate through the medium facing energy loss and diffusion. To describe this process one has to solve the transport equation, that describes how the particles propagate in the magnetic field, losing and/or gaining energy.  For a stationary solution, in the limit of spherical symmetry and neglecting advective and convective transport terms, it has the form~\cite{Strong_Moskalenko}:

\begin{equation}
    -\frac{1}{r^2}\left [r^2 D \frac{\partial f}{\partial r}\right ]+\frac{1}{p^2}\frac{\partial}{\partial p} (\dot{p} p^2 f) = q(r,p),
    \label{eq:diffusion}
\end{equation}

\noindent where $f(r,p)$ is the  $e^\pm$ distribution function at equilibrium at radius $r$ and momentum $p$, related to the energy distribution at energy $E$ by $\frac{dn_e}{dE}(r,E)=4\pi p E f(r,p)$, while the source term $q(r,p)$ is related to that of Eq.~(\ref{eq:DM_source_fn}) by $Q_e(r,E)=4\pi p E q(r,p)$. In the equation above the first term on the left--hand side describes the spatial diffusion, with $D(r,p)$ the diffusion coefficient, and the second term describes the energy loss due to radiative processes. When diffusion is neglected (i.e. for $D(r,p)=0$) Eq.~(\ref{eq:diffusion}) has explicit solution~\cite{multifrequency_coma_cluster_2006, Regis_Ullio_multiwave_2007, radio_signal_M87_2015}: 

\begin{equation} 
\frac{dn_e}{dE} (r,E) = \frac{1}{b(r,E)} \int_E^{m_{\chi}} dE' \, Q_e (E',r) \, .
\label{eq:dndE}
\end{equation}
The term $b(r, E)=-dE/dt$ corresponds to the total energy loss rate. For energetic $e^\pm$'s it is dominated by synchrotron emission in the magnetic field close to the BH and Inverse Compton (IC) scattering events off the CMB background along the line--of--sight from the source to the observer: 
\begin{equation} 
b(r,E) = b_{\rm syn}(r,E) + b_{\rm IC}(r,E) \, ,
\label{eq:b_E}
\end{equation}
where 
\begin{equation} 
b_{\rm syn}(r,E) = \frac{4}{3} \sigma_{\rm T} c \frac{B(r)^2}{2 \mu_0} \left(\frac{E}{m_e c^2}\right)^2
\label{eq:b_E_syn}
\end{equation}
\begin{equation} 
b_{\rm IC}(r,E) = \frac{4}{3} \frac{\sigma_{\rm T}}{m^2_e} E^2 u_{\rm CMB}
\label{eq:b_E_IC}
\end{equation}
with $\sigma_{\rm T}$ the Thomson cross-section, $B(r)$ the intensity of the magnetic field, 
$c$ the speed of light, $\mu_0$ the vacuum permeability and $u_{\rm CMB} \simeq 0.26$ $\rm eV \, cm^{-3}$ 
the background CMB energy density \cite{Aloisio:2004hy, multifrequency_coma_cluster_2006, Regis_Ullio_multiwave_2007, Cirelli:2010xx, radio_signal_M87_2015}. In particular the two energy--loss mechanisms are anti-correlated,  since only the electrons that do not lose energy close to the BH due to synchrotron emission can scatter off the CMB photons far way from the BH to generate the IC signal.

\begin{figure*}[ht!]
\centering
\includegraphics[width=7.49cm,height=6cm]{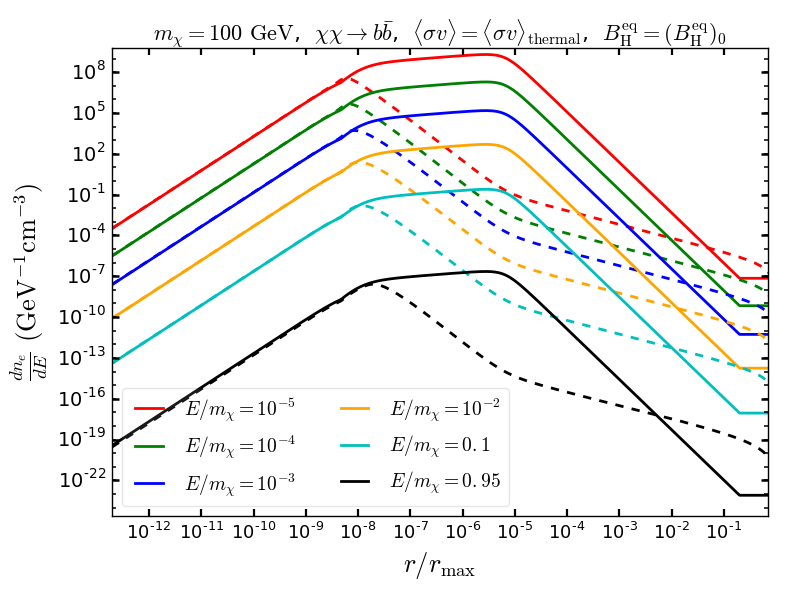}
\includegraphics[width=7.49cm,height=6cm]{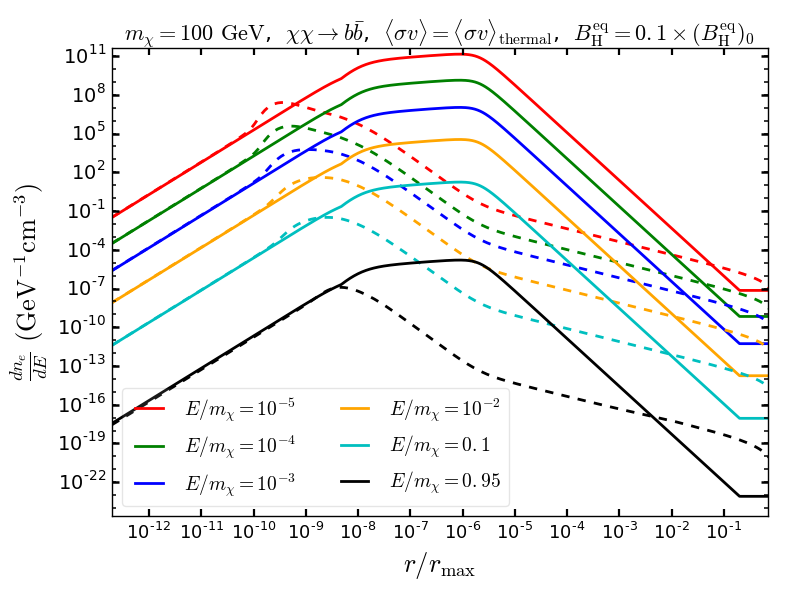}
\caption{The solution ${dn_e}/{dE}$ of Eq.~\eqref{eq:diffusion} as a function of the radius $r$ for 
different values of the energy $E$, neglecting (solid lines) and considering (dashed lines) diffusion. 
The value of $r_{\rm max}$ is assumed to be equal to $r_{\rm in}$ (defined in Eq.~\eqref{eq:r_in}). 
Left: $B_{\rm H}^{\rm eq}$ = $(B_{\rm H}^{\rm eq})_0$; right: $B_{\rm H}^{\rm eq}$ = $0.1\times(B_{\rm H}^{\rm eq})_0$ (with $(B_{\rm H}^{\rm eq})_0$ defined in Eq.~(\ref{eq:B_eq_H})).}
\label{fig:dndE_solution}
\end{figure*}

In the case of the SMBH's in the Galactic Center~\cite{Regis_Ullio_multiwave_2007} or in M87~\cite{radio_signal_M87_2015} the effect of diffusion is usually negligible, since the distance scale over which the $e^\pm$ lose most of their energy is small compared to the size of the system, so that Eq~(\ref{eq:dndE}) can be directly used to calculate signals. However LMBX's are much smaller systems so, especially in the case of heavy WIMPs producing high--energy final states, diffusion can be non--negligible. This is shown in Fig.~\ref{fig:dndE_solution}, where for $m_\chi$ = 100 GeV, $\langle \sigma v \rangle$ = $\langle \sigma v \rangle_{\rm thermal}$ and $\chi \chi \rightarrow b\bar{b}$ the solid lines show $\frac{dn_e}{dE}$ calculated using Eq.~(\ref{eq:dndE}) without diffusion as a function of the distance $r$ from the center for different emission energies, while the dashed lines indicate the same quantity when the effect of diffusion is included. To calculate the latter curves we have solved Eq.~\eqref{eq:diffusion} numerically with boundary conditions $\frac{dn_e}{dE}(r=r_{\rm in})$ = 0, 
$\frac{\partial}{\partial r}\left(\frac{dn_e}{dE}\right)(r=r_{\rm in})$ = 0 (with $r_{\rm in}$ the influence radius defined in Section~\ref{sec:spike_profile}) and $\frac{dn_e}{dE}(E=m_\chi)$ =0~\footnote{We used the \href{https://reference.wolfram.com/language/tutorial/NDSolveMethodOfLines.html}{``Method of Lines"} in the $\texttt{Mathematica}$ function \href{https://reference.wolfram.com/language/ref/NDSolve.html}{\texttt{NDSolve}}.}. Diffusion is connected with unknown variables needed in the description of turbulence such as the amplitude of the random magnetic field and the scale of the turbulence spectrum. For the diffusion parameter we adopt $D(r,p)= (1/3) r_g v_e$, 
with $r_g = E/(e B(r))$  the gyroradius of the electron and $v_e$ the electron velocity. This choice corresponds to Bohm diffusion, when the coherence length of the magnetic field is comparable or greater that $r_g$~\cite{Bohm_Regis_2023}.

From the left--hand plot of Fig.~\ref{fig:dndE_solution} one can see that, for 
$B_{\rm H}$ = $(B_{\rm H}^{\rm eq})_0$  (see Eq.~(\ref{eq:B_eq_H})) the effect of diffusion is to displace the $e^\pm$ density to outer radii for $r\gtrsim 10^{-8} r_{\rm in}$. In Fig.~\ref{fig:rho_r} one can notice that for $\langle \sigma v\rangle$ = $\langle \sigma v\rangle_{\rm thermal}$ the annihilation plateau extends to approximately the same distance from the origin, i.e. for $r\lesssim 10^{-8} r_{\rm in}$. In this case one can expect a sizeable reduction of the expected signals compared to the case when diffusion is neglected, since the two effects of diffusion and of the annihilation plateau are complementary over the full range of $r$. However, in Section~\ref{sec:results} we will also see that, unless $m_\chi$ reaches the TeV scale, the present observational data on BH–LMXB's imply bounds on $\langle \sigma v\rangle$ that are several orders of magnitude smaller than $\langle \sigma v\rangle_{\rm thermal}$. In such case the extension of the annihilation plateau is smaller and the inner part of the $e^\pm$ spectrum, not affected by diffusion, extends deep into the DM spike where the density is enhanced, mitigating the effect of diffusion. The right--hand plot of Fig.~\ref{fig:dndE_solution} shows instead the $e^\pm$ spectrum for $B_{\rm H}$ = $0.1 \times (B_{\rm H}^{\rm eq})_0$. In this case one can see that diffusion suppresses the $e^\pm$ density over a wider range of $r$ that extends closer to the center compared to when $B_{\rm H}$ = $(B_{\rm H}^{\rm eq})_0$. Indeed, diffusion can be neglected as long as the $e^\pm$'s can be assumed to lose their energy instantaneously upon production, and this is sensitive to the intensity of the magnetic field.  For this reason the numerical analysis of Sec.~\ref{sec:results}  will allow to conclude that the main effect of diffusion is rather mild as long as $B_{\rm H}$ = $(B_{\rm H}^{\rm eq})_0$, while diffusion is instrumental in enhancing the sensitivity of the results upon the value of $B_{\rm H}$.  

\subsubsection{Synchrotron signal}
\label{eq:synchrotron}
The synchrotron emissivity generated by the $e^\pm$'s produced from DM annihilation 
is given by \cite{multifrequency_coma_cluster_2006, radio_signal_M87_2015}: 
\begin{equation} 
j_{\rm syn}(\nu, r) = 2 \int^{m_\chi}_{m_e} dE \, \frac{dn_e}{dE} (r,E) \, P_{\rm syn}(\nu, r, E) \, , 
\label{eq:j_syn}
\end{equation}

\noindent where the synchrotron power is \cite{Aloisio:2004hy, multifrequency_coma_cluster_2006, radio_signal_M87_2015}: 
\begin{equation} 
P_{\rm syn}(\nu, r, E) = \frac{1}{4 \pi \varepsilon_0} \frac{\sqrt{3} e^3 B(r)}{m_e c} 
G_i \left(\frac{\nu}{\nu_c(r,E)}\right) \,.
\label{eq:P_syn}
\end{equation}
Here $e$ is the elementary charge, $\varepsilon_0$ is the vacuum permittivity, 
$\nu_c(r,E)$ is the critical frequency given by:
\begin{equation} 
\nu_c(r,E) = \frac{3 e E^2 B(r)}{4 \pi m^3_e c^4} \, ,
\label{eq:nu_c}
\end{equation}
and the function $G_i$ is the isotropic synchrotron spectrum: 
\begin{equation} 
G_i (x) = \frac{1}{2} \int^{\pi}_0 G\left( \frac{x}{\sin \alpha} \right) \sin^2\alpha \, d\alpha
\label{eq:Gi_fn}
\end{equation}
with $G(t) = t \int^\infty_t \, K_{5/3} (u) du$, where $K_{5/3}$ is the modified Bessel 
function of order 5/3. 

\noindent Finally, the synchrotron signal at observation \cite{multifrequency_coma_cluster_2006, radio_signal_M87_2015} is:
\begin{equation}
S_{\rm syn}(\nu) = \frac{1}{4\pi} \int_{\Delta \Omega} d\Omega \int_{l.o.s.} dl \hspace{1mm} 
j_{\rm syn} (\nu, l, \theta) \, ,
\label{eq:syn_signal}
\end{equation}
where the line-of-sight (l.o.s.) coordinate $l$ is related to the radial coordinate $r$ 
and the angle $\theta$ (with respect to the direction of the BH) as 
$r = \sqrt{d^2 + l^2 - 2 \, l \, d \, \cos{\theta}}$, with $d$ being the distance of the BH from 
Earth (for XTE J1118+480, $d \simeq 1.7$ kpc). 
The solid angle $\Delta \Omega$ is the field-of-view of interest. 
%Since $\frac{dn_e}{dE}$ or $j_{\rm syn}$ is concentrated in a region 
%where the spike in $\rho_{\rm DM}$ 

Note that at low radio frequencies where the synchrotron signal dominates 
the effect of synchrotron self-absorption (SSA) may become relevant. In this process the emitted synchrotron radiation can be re-absorbed by the radiating $e^\pm$ along the line-of-sight. The effect of the SSA can be quantified in terms of the opacity to the absorption process at a given frequency $\nu$, namely, by the quantity $\tau_\nu (\theta)$, where $\theta$ is the angle of the line-of-sight with respect to the direction of the BH (see \cite{Aloisio:2004hy, Regis_Ullio_multiwave_2007} for the details). As explained in \cite{Aloisio:2004hy}, the effect of SSA becomes important only when $\tau_\nu$ approaches or exceeds unity and it can be neglected when $\tau_\nu \ll 1$. For our BH system we have checked that for the frequency range used here the parameter $\tau_\nu$ is always much smaller than unity ($\lesssim 10^{-2}$). Hence, we can ignore the effect of SSA in our analysis.

\subsubsection{Inverse Compton (IC) signal}
\label{sec:inverse_compton}
Inverse Compton (IC) scattering events of the DM induced $e^\pm$'s on the CMB photons give rise to a photon spectrum that ranges from below the extreme ultra-violet frequency range up to the soft gamma-ray band, and that peaks usually in the soft X-ray energy band. The IC power 
produced in the scattering of one $e^-(e^+)$ with the CMB is given by: 
\begin{equation} 
P_{\rm IC} (E_\gamma, E) = c \, E_\gamma \int d\epsilon \,\, n(\epsilon) \, 
\sigma(E_\gamma, \epsilon, E) .
\label{eq:P_IC}
\end{equation}
Here $E$ and $E_\gamma$ are, respectively, the energies of the $e^-(e^+)$ and of the produced photon, $n(\epsilon)$ is the differential number density of the CMB photons (of energy $\epsilon$) and $\sigma(E_\gamma, \epsilon, E)$ is the IC scattering cross-section (see \cite{multifrequency_coma_cluster_2006, Cirelli:2010xx} for details). Then the observed IC signal is:
\begin{equation}
S_{\rm IC}(E_\gamma) = \frac{1}{4\pi} \int_{\Delta \Omega} d\Omega \int_{l.o.s.} dl \hspace{1mm} 
j_{\rm IC} (E_\gamma, l, \theta),
\label{eq:IC_signal}
\end{equation}
\noindent with the emissivity of IC photons given by: 
\begin{equation} 
j_{\rm IC}(E_\gamma, r) = 2 \int^{m_\chi}_{m_e} dE \, \frac{dn_e}{dE} (r,E) \, 
P_{\rm IC}(E_\gamma, E) \,.
\label{eq:j_IC}
\end{equation}
\noindent In Eq.~(\ref{eq:IC_signal}) $\Delta \Omega$ is the detector angular field-of-view.

\subsection{Radio and X-ray data from XTE J1118+480}
\label{sec:observations}
In Ref.~\cite{radio_data_XTE} coordinated radio and X-ray observations of 
XTE J1118+480 in quiescence are reported. In particular, the radio observation, 
performed with the Karl G. Jansky Very Large Array, corresponds to an emission
of 2.5 $\pm$ 0.74$\times 10^{-19}$ erg/cm$^2$/s integrated over the frequency range $\nu\le$ 5.3 GHz
with the detector angular field-of-view 
$\Delta\Omega$ = $\Delta \Omega_{\rm radio} \simeq$ 1.17$\times$10$^{-9}$ sr.
On the other hand, the X-ray emission corresponds
to 1.2$\times$10$^{-14}$ erg/s/cm$^2$ integrated in the frequency range 2.4$\times 10^{17}$ Hz $\le\nu\le$ 2.4$\times10^{18}$ Hz, measured by the Chandra X-ray Telescope in the angular field-of-view 
$\Delta\Omega$ = $\Delta \Omega_{\rm X-ray} \simeq$ 1.8$\times$10$^{-11}$ sr~\cite{Chandra_Weisskopf:2003bam}. Both measurements are reported in Fig~\ref{fig:flux_vs_freq}, and will be used in the next Section as upper limits on the signals 
to put constraints on the WIMP annihilation cross section times velocity $\langle \sigma v \rangle$. In particular, using the synchrotron and IC emissivities defined in Eqs.~\eqref{eq:j_syn} and \eqref{eq:j_IC} we estimate the frequency distributions of the DM induced radio and X-ray signals at observation as: 
\begin{equation}
S_{\rm radio}(\nu) = \frac{1}{4\pi} \int_{\Delta \Omega_{\rm radio}} d\Omega \int_{l.o.s.} dl \hspace{1mm} 
\left[j_{\rm syn} (\nu, l, \theta) + j_{\rm IC} (\nu, l, \theta) \right] \, ,
\label{eq:S_radio}
\end{equation}

\begin{equation}
S_{\rm X-ray}(\nu) = \frac{1}{4\pi} \int_{\Delta \Omega_{\rm X-ray}} d\Omega \int_{l.o.s.} dl \hspace{1mm} 
\left[j_{\rm syn} (\nu, l, \theta) + j_{\rm IC} (\nu, l, \theta) \right] \, ,
\label{eq:S_X_ray}
\end{equation}
and put constraints on $\langle \sigma v\rangle$ by comparing the observed integrated radio and X-ray luminosities (in units of erg/cm$^2$/s):
\begin{equation}
L_{\rm radio} = \int^{5.3 \, \rm GHz}_{\nu = 1 \, \rm GHz} d\nu \, S_{\rm radio}(\nu) \, ,
%\label{}
\end{equation}

\begin{equation}
L_{\rm X-ray} = \int^{2.4\times10^{18} \, \rm Hz}_{\nu = 2.4\times10^{17} \, \rm Hz} d\nu \, 
S_{\rm X-ray}(\nu) \, ,
%\label{}
\end{equation}
to the corresponding experimental upper bounds.

\section{Results}
\label{sec:results}

\begin{figure*}[ht!]
\centering
\includegraphics[width=7.49cm,height=6cm]{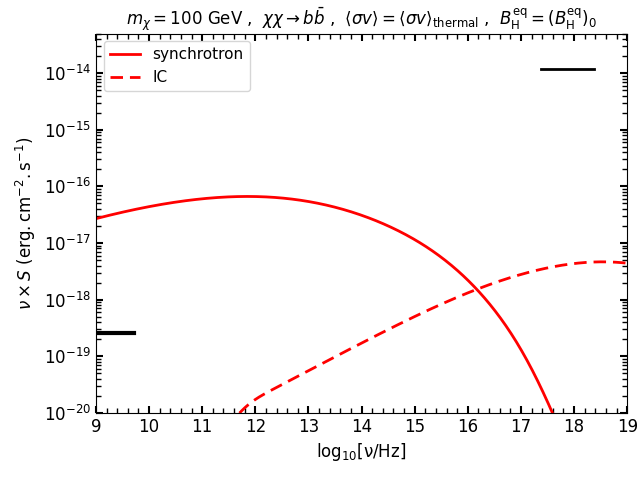}
\caption{Frequency distributions of the DM induced signals from XTE J1118+480 
BH-binary. The red solid and dashed lines show the synchrotron and the IC signals, i.e., $S_{\rm syn}$ (Eq.~\eqref{eq:syn_signal}) and $S_{\rm IC}$ (Eq.~\eqref{eq:IC_signal}), respectively, multiplied by the frequency. Here for $S_{\rm syn}$ and $S_{\rm IC}$ the corresponding field-of-views ($\Delta \Omega$'s) are taken to be $\Delta \Omega_{\rm radio}$ and 
$\Delta \Omega_{\rm X-ray}$, respectively. 
The two bars represent the radio and the X-ray flux observed from the direction of XTE J1118+480, integrated over the corresponding observational frequency ranges. The bar levels indicate the values of the integrated observed flux while their horizontal lengths indicate the corresponding frequency ranges over which the flux is integrated. In both curves the magnetic field at the event horizon is fixed to the equipartition value.}
\label{fig:flux_vs_freq}
\end{figure*}

In this Section we will obtain our results fixing the magnetic field at the event horizon $B^{\rm eq}_{\rm H}$ to the equipartition value given by Eq.~(\ref{eq:B_eq_H}). Smaller values of the magnetic field are discussed in Section~\ref{app:b_sens}.

In Fig.~\ref{fig:flux_vs_freq} the expected signals from WIMP annihilation in the two channels of synchrotron radiation and Inverse Compton (calculated neglecting diffusion and following the procedures outlined, respectively, in Section~\ref{eq:synchrotron} and Section~\ref{sec:inverse_compton}) are compared to the experimental observations summarized in Section~\ref{sec:observations} for $m_\chi$ = 100 GeV,  $\langle \sigma v\rangle$ = $\langle \sigma v\rangle_{\rm thermal}$ and for the $\chi\chi\rightarrow b\bar{b}$ annihilation channel. From such specific example one can observe that the synchrotron signal exceeds the experimental data in the radio frequency range by several orders of magnitude and so is potentially able to put interesting constraints on $\langle \sigma v\rangle$.
On the other hand, in the X--ray part of the spectrum the Inverse Compton expected signal is much smaller than the level of observations. This feature turns out to be common to the full range of WIMP masses that we consider in our analysis, so all the bounds discussed in this Section will be driven by the synchrotron signal.

\begin{figure*}[ht!]
\centering
\includegraphics[width=16cm,height=5.5cm]{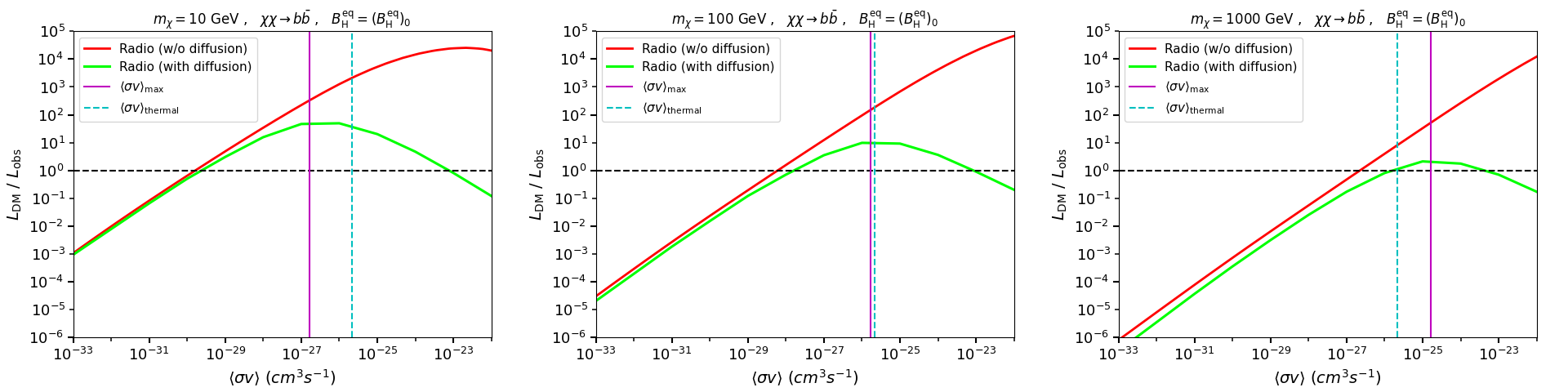}
\caption{DM induced radio luminosities (normalised to the measured one) as a function of 
$\langle \sigma v\rangle$ are shown for three values of the DM mass, $m_\chi = 10$ GeV (left), 100 GeV (middle) 
and 1000 GeV (right), considering $B_{\rm H}^{\rm eq}$ = $(B_{\rm H}^{\rm eq})_0$. The green solid lines include the effect of diffusion while the red solid lines neglect it. The vertical purple line in each 
panel indicates the maximum value of $\langle \sigma v\rangle$ beyond which the annihilation plateau in the DM density reaches the location of the 
companion star in the BH-binary. The vertical cyan dashed line in each plot represents the standard value of $\langle \sigma v\rangle$ for a thermal WIMP.}
\label{fig:Lumi_vs_sv}
\end{figure*}

The effect of diffusion is shown in Fig.~\ref{fig:Lumi_vs_sv}, where the expected synchrotron signal normalized to the bound is plotted as a function of $\langle \sigma v\rangle$ for $m_\chi$ = 10, 100, 1000 GeV. In each plot the green solid line includes the effect of diffusion and the red solid line neglects it. In both cases as long as $\langle \sigma v\rangle$  is small enough and the density plateau due to annihilation (see Fig.~\ref{fig:rho_r}) is negligible the signal grows linearly with $\langle \sigma v\rangle$.  On the other hand higher values of $\langle \sigma v\rangle$ imply a larger annihilation plateau that erases the overdensity of the cusp (in particular, neglecting diffusion the signal is proportional to $\rho_{\rm sat}^2\times \langle \sigma v\rangle \propto \langle \sigma v\rangle^{-2}\times \langle \sigma v\rangle$ and so it scales as $1/\langle \sigma v\rangle$). From Fig.~\ref{fig:Lumi_vs_sv} one can see that diffusion suppresses the signal sizeably when $\langle \sigma v\rangle\simeq\langle \sigma v\rangle_{\rm thermal}$  but has a more moderate effect when  $\langle \sigma v\rangle$ hits the experimental bound for $\langle \sigma v\rangle<<\langle \sigma v\rangle_{\rm thermal}$. As already pointed out in Section~\ref{sec:synchrotron_and_IC} this can be understood noticing that when $\langle \sigma v\rangle$ is small enough the annihilation plateau extends to smaller radii and the $e^\pm$ spectrum that drives the signal extends deep into the DM spike where it is not affected by diffusion (see Fig.~\ref{fig:dndE_solution}). The narrow interval of the spike index $\gamma_{\rm sp}$ = 1.85$_{-0.04}^{+0.04}$ for XTE J1118+480 that we use in our analysis was obtained the Ref.~\cite{DM_spike_LMBH_2023} assuming that the companion star in the BH-LMXB system orbits inside the spike, i.e. that its orbital radius is larger than that of the annihilation plateau. This implies an upper bound $\langle \sigma v \rangle_{\rm max}$, which in all the plots of Fig.~\ref{fig:Lumi_vs_sv} is represented by the vertical purple line. Figure~\ref{fig:Lumi_vs_sv} shows that the results of our analysis are consistent with such assumption, since the signals hit the bound for $\langle \sigma v \rangle<\langle \sigma v \rangle_{\rm max}$.

\begin{figure*}[ht!]
\centering
\includegraphics[width=14cm,height=10.5cm]{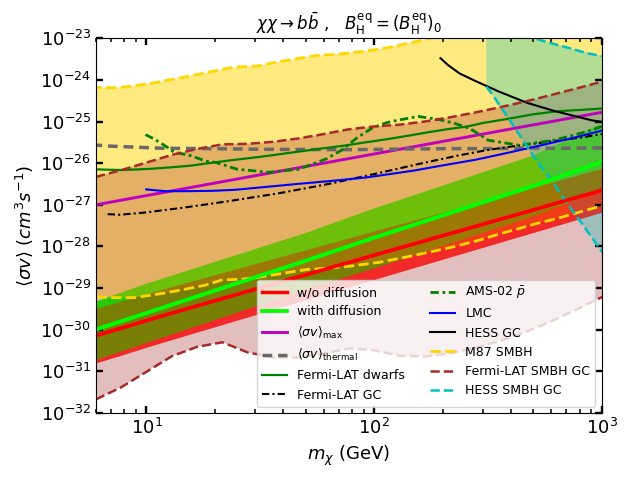}
\caption{Upper bounds on $\langle \sigma v\rangle$ as a function of $m_\chi$. The green and the red solid curves represent the bounds from XTE J1118+480 calculated in the present paper when the effect of spatial diffusion is included and neglected, respectively, and assuming the central value for the determination of the spike index $\gamma_{\rm sp} = 1.85$ that explains its abnormally fast orbital decay. On the other hand the green and the red bands show the corresponding variation of the limits when the 1$\sigma$ uncertainty in $\gamma_{\rm sp}$ is considered. The purple line indicates the maximum $\langle \sigma v\rangle$ that ensures that, for consistency with the analysis of Ref.~\cite{DM_spike_LMBH_2023}, the companion star orbit of XTE J1118+480 is outside the annihilation plateau. The other constraints correspond to Fermi-LAT dwarf galaxies~\cite{Fermi-LAT_dSph:2016uux} (solid dark green), Fermi-LAT GC~\cite{Abazajian:2020tww} (black dash-dotted), 
AMS anti-proton~\cite{Calore:2022stf} (dark green dash-dotted), EMU LMC (solid blue)~\cite{LMC_Regis:2021glv}, HESS GC (solid black)~\cite{HESS_GC:2022ygk}, M87 SMBH (yellow dashes)~\cite{radio_signal_M87_2015}, 
Fermi-LAT SMBH GC (brown dashes)~\cite{Fields_gamma_excess_2014} and HESS SMBH GC (cyan dashes)~\cite{HESS_SMBH_Balaji:2023hmy}. Each curve is surrounded by a band of the same color that represents the uncertainty of the bound upon the DM spike index. The gray dashed line represents the thermal value of 
$\langle \sigma v\rangle$ corresponding to the observed DM relic abundance for a thermal WIMP~\cite{Thermal_relic_Steigman:2012nb}.}
\label{fig:sv_mx}
\end{figure*}

The final result of our analysis is shown in Fig.~\ref{fig:sv_mx}, where the upper bound on $\langle \sigma v\rangle$ as a function of $m_\chi$ obtained from XTE J1118+480 is compared to the existing constraints on the same quantity from other observations. In particular the red solid line represents the upper bound on $\langle \sigma v\rangle$ when the effect of diffusion is neglected, and the red band represents how the same bound changes when the uncertainty of the determination of the spike index $\gamma_{\rm sp}$ in Ref.~\cite{DM_spike_LMBH_2023} is considered. In the same Figure the green solid line and the green band represent the same quantities when diffusion is taken into account. One can see from such Figure that XTE J1118+480 allows to constrain $\langle \sigma v\rangle$ below $\langle \sigma v\rangle_{\rm thermal}$ in the full range of $m_\chi$ also when the uncertainty on $\gamma_{\rm sp}$ is included. 

In Figure~\ref{fig:sv_mx} the bound from XTE J1118+480 is compared to those obtained from the super massive BHs (SMBHs) in the Galactic Center (GC)~\cite{Fields_gamma_excess_2014, HESS_SMBH_Balaji:2023hmy} or in M87~\cite{radio_signal_M87_2015}, including the uncertainties due to the corresponding spike indices. We directly report the bounds corresponding to the observation of the SMBH in M87 and the H.E.S.S. observation of the SMBH in the GC from \cite{radio_signal_M87_2015} and \cite{HESS_SMBH_Balaji:2023hmy}, respectively. On the other hand, we obtain the bound corresponding to the Fermi-LAT observation of the SMBH in the GC by using the methodology and observational data provided in~\cite{Fields_gamma_excess_2014} as well as the range 1.5 $\leq \gamma_{\rm sp} \leq$ 2.36 for the spike index. All such bounds (represented by 
yellow, cyan and brown bands, respectively) do not reach $\langle\sigma v\rangle_{\rm thermal}$ in the full range of $m_\chi$ (with the exception of the Fermi-LAT observation of the SMBH in the GC for $m_\chi \lesssim 15$ GeV) when the uncertainties on the corresponding spike indices (much larger than that obtained for XTE J1118+480) are included. In particular, for the annihilation channel that we considered ($\chi\chi \rightarrow b\bar{b}$) up to $\simeq$ 1 TeV WIMP masses the bound from XTE J1118+480 is the most constraining compared to all others, including the gamma-ray signals from dwarf galaxies~\cite{Fermi-LAT_dSph:2016uux} and GC~\cite{Abazajian:2020tww}, the radio signal from LMC~\cite{LMC_Regis:2021glv} and the anti-particle flux measured by AMS~\cite{Calore:2022stf}. 

Notice, however, that, as already pointed out,  the results of this Section rest on the assumption that the magnetic field corresponds to the equipartition value. 

\subsection{Sensitivity to the magnetic field and comparison with observation}
\label{app:b_sens}

\begin{figure*}[ht!]
\centering
\includegraphics[width=7.49cm,height=6cm]{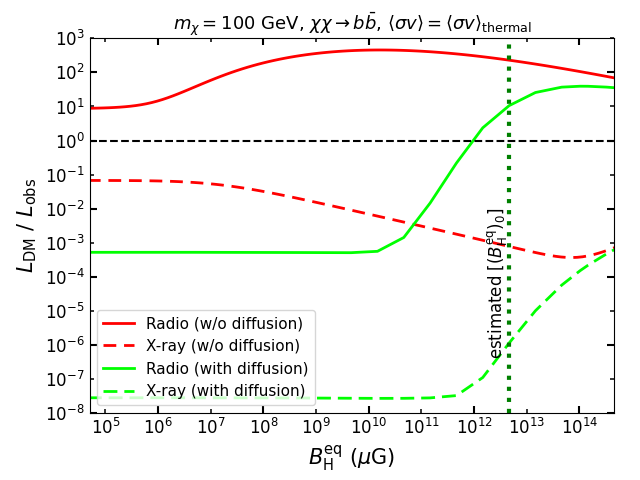}
\caption{Dependence of the radio and X-ray signal luminosities (normalised to the measured ones) 
on the value of $B_{\rm H}^{\rm eq}$. 
The red (solid and dashed) curves show the results neglecting diffusion, while the green (solid and dashed) 
curves including it. The solid (red and green) curves are corresponding to the radio signal, while the dashed (red and green) curves are for the X-ray. The vertical green dotted line indicates the equipartition value $(B_{\rm H}^{\rm eq})_0$ of magnetic field at the event horizon quoted in Section~\ref{sec:magnetic_field}.}
\label{fig:flux_vs_BH}
\end{figure*}

The results shown in Fig~\ref{fig:sv_mx} assume $B_{\rm H}^{\rm eq}$ = $(B_{\rm H}^{\rm eq})_0$ for the magnetic field at the event horizon $B_{\rm H}^{\rm eq}$, with $(B_{\rm H}^{\rm eq})_0$ related to the accretion rate of the BH by the equipartition formula of Eq.~(\ref{eq:B_eq_H}). One may ask how robust is such assumption and how sensitive is the radio synchrotron signal on the value of $B_{\rm H}^{\rm eq}$. 

On general grounds~\cite{equipartition1, equipartition2, Frolov_2010, piotrovich_2010} the equipartition assumption is made on the basis of scaling and energetics
arguments. In particular, in a binary system the magnetic energy density grows faster than the kinetic energy, reaching equipartition quickly, and maintaining it relies on the existence of dissipation mechanisms in the disk to keep the magnetic field at its equipartition value. The ensuing 
characteristic scale of the magnetic field is expected to be of the order of $10^{10}$ $\mu$G near the horizon of a supermassive black hole ($M_{\rm BH}\simeq 10^9 M_\odot$)  and of the order $10^{14}$ $\mu$G near the horizon of a stellar mass one (as in XTE J1118+480), with a theoretical upper bound provided by the Eddington limit on the accretion rate~\cite{gamma_obsv_FermiLAT_2017}, corresponding to $\dot{m}$ =1 in Eq.~(\ref{eq:B_eq_H}). 

In Fig.~\ref{fig:flux_vs_BH} the expected radio and X-ray signals, normalized to the respective bounds, are plotted as a function of the $B_{\rm H}^{\rm eq}$ parameter and for the same choice for the other parameters of Fig.~\ref{fig:flux_vs_freq}. In particular, in Figure~\ref{fig:flux_vs_BH} the red (green) solid line indicates the radio synchrotron signal prediction when the effect of diffusion is neglected (included).
%while the red (green) dashed line shows the IC signal with (without) the inclusion of diffusion. 
In Figure~\ref{fig:flux_vs_BH} the vertical green dotted line corresponds to the equipartition value $B_{\rm H}^{\rm eq}$ = $(B_{\rm H}^{\rm eq})_0$. From this plot one can see that when the effect of diffusion is neglected and as long as synchrotron emission dominates, when $B_{\rm H}^{\rm eq}<(B_{\rm H}^{\rm eq})_0$ the signal not only does not drop but actually {\it increases}. In fact it is proportional to $dn_e/dE\times P_{\rm syn}$ (see Eq.~(\ref{eq:j_syn})) with $dn_e/dE\propto 1/B^2$ when in the denominator of Eq.~(\ref{eq:dndE}) one has $b\simeq b_{\rm syn}$ (see Eqs.~(\ref{eq:b_E}) and~(\ref{eq:b_E_syn})), while in Eq.~(\ref{eq:j_syn}) the quantity 
$P_{\rm syn}$ is only linear in $B$. We conclude that when diffusion is neglected the bounds obtained using the equipartition value for $B_{\rm H}^{\rm eq}$ are robust. However, the same figure shows that when diffusion is included the signal becomes more sensitive to $B_{\rm H}^{\rm eq}$. In this case, while for $B_{\rm H}^{\rm eq}$ = $(B_{\rm H}^{\rm eq})_0$ the effect of diffusion on our results is mild, diffusion is instrumental in enhancing the sensitivity on the value of the magnetic field of the signal predictions and, as a consequence, of the bounds. 

\begin{figure*}[ht!]
\centering
\includegraphics[width=7.49cm,height=6cm]{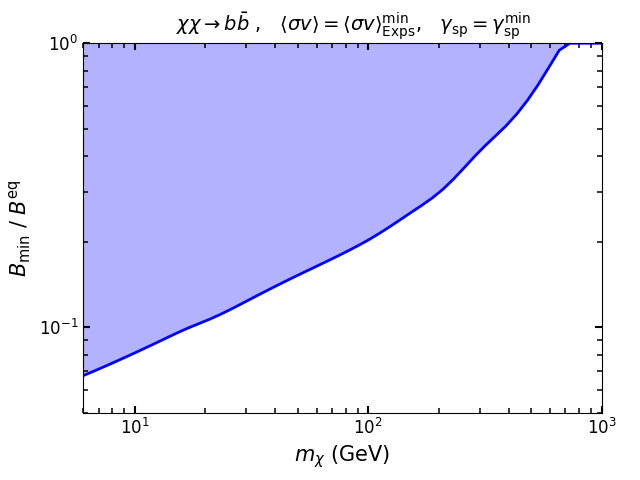}
\caption{Minimal value of the magnetic field $B$ normalized to the equipartition value $B^{\rm eq}$ required for the bound on $\langle \sigma v\rangle$ from the XTE J1118+480 binary to be more constraining than the combination $\langle \sigma v\rangle^{\rm min}_{\rm Exps}$ of the existing bounds  from other sources. The effect of diffusion is included in the calculation of the expected signal. The spike index $\gamma_{\rm sp}$ is fixed to the lower edge of the determination of Ref.~\cite{DM_spike_LMBH_2023}.}
\label{fig:Bmin}
\end{figure*}

Indeed, the recent study~\cite{dallilar_2018} of V404
Cygni~\footnote{We thank the anonymous referee for pointing this measurement out to us.}, a black hole system quite similar to XTE J1118+480, leads to an estimation of the magnetic field about a factor of 20 smaller than its equipartition value ($\simeq$ 33 G vs. $\simeq$650 G). Such estimation is much lower than previous estimates from other sources and, as pointed out by the authors, requires an electron energy distribution with Lorentz factors higher than usually assumed, with a corresponding process to accelerate the electrons that is currently unclear. However, it is not possible to rule out the possibility that it is representative of the possible excursion of the magnetic field close to a stellar--mass BH. It then becomes relevant to assess by how much the magnetic field of XTE J1118+480 can be reduced compared to the equipartition value to preserve the results shown in Fig.~\ref{fig:sv_mx}. This is done in Fig.~\ref{fig:Bmin}, that shows as a function of the WIMP mass $m_\chi$ the minimal reduction factor for the magnetic field compared to the equipartition value that is required for the bound on $\langle \sigma v\rangle$ from the XTE J1118+480 binary to be more constraining than the combination $\langle \sigma v\rangle^{\rm min}_{\rm Exps}$ of the existing bounds from other sources (namely gamma observation from Fermi-LAT at the GC and radio observation from LMC, see Fig.~\ref{fig:sv_mx}). One can see from this plot that, in order to relax the XTE J1118+480 limit to the level of other existing bounds a reduction factor of about 14 at low $m_\chi$, becoming progressively smaller at higher WIMP masses (and eventually saturating to one) is required. This plot implies that, as far as the V404 Cygni measurement of the magnetic field is concerned, a reduction factor of 20 would be too large to keep the XTE J1118+480 constraint competitive with other observations (in particular we checked that it would correspond to a DM signal from XTE J1118+480 below observation in the full $m_\chi$ range). We conclude that the determination of the intensity of the magnetic field in BH–LMXB systems represents the major uncertainty in using them to search for WIMPs as an alternative to heavier BHs.

\section{Conclusions}
\label{sec:conclusions}

Although alternative scenarios have been put forward~\cite{justham_2006, Chen_Lee_2015} it has been recently claimed that the dynamical friction between DM and the companion star orbiting around the low--mass Black Hole ($\simeq$ a few $M_\odot$) of two X–ray binaries, A0620-00 and XTE J1118+480, can satisfactorily explain their abnormally fast orbital decays~\cite{DM_spike_LMBH_2023}. Assuming this is so, by comparing the prediction of the orbital decay of the X–ray binaries to the observational value it is possible to pin down the index $\gamma_{\rm sp}$ of the DM spikes surrounding their black holes with an accuracy of $\simeq$ a few percent, way better than that for much bigger systems such as the super massive BHs in the Galactic Center or in M87. This is a crucial piece of information to reduce the uncertainties on the bounds that can be obtained from such systems, since the calculation of DM annihilation signals is very sensitive to $\gamma_{\rm sp}$.

In the present paper we have used data from XTE J1118+480, which among the
two binaries has the largest spike index, to put bounds on the WIMP annihilation cross section times velocity $\langle \sigma v\rangle$, assuming 
that DM annihilation is driven by the $\chi\chi\rightarrow b\bar{b}$ annihilation channel and that it proceeds in $s$--wave. The bounds are driven by the radio synchrotron signal produced by $e^\pm$ final states propagating in the magnetic field in the vicinity of the BH, while the Inverse Compton scattering events induced by $e^\pm$ on CMB photons generate an X--ray signal which is always subdominant. To be consistent with the analysis of Ref.~\cite{DM_spike_LMBH_2023} we assumed that the BH in the binary and the magnetic field $B$ are spherically symmetric. Moreover, we adopted as a reference value for the magnetic field at the event horizon $B_{\rm H}$ the intensity
that corresponds to assuming equipartition between the
magnetic and kinetic energy densities (see Eq.~(\ref{eq:B_eq_H})).

With such assumptions, we find that XTE J1118+480 allows to constrain $\langle \sigma v\rangle$ well below its thermal value $\langle \sigma v\rangle_{\rm thermal}$ (corresponding to the observed DM relic density in the Universe for a standard thermal WIMP) in the full range of $m_\chi$, also when the uncertainty on the DM spike index $\gamma_{\rm sp}$ is included. This is at variance with the bounds from the observations of the SMBHs in the GC and in M87, which, in spite of potentially producing much larger DM densities compared to XTE J1118+480,  do not reach $\langle\sigma v\rangle_{\rm thermal}$ in the full range of $m_\chi$ (with the exception of the Fermi-LAT observation of the SMBH in the GC for $m_\chi \lesssim 15$ GeV) when the very large uncertainties on the corresponding spike indices are taken into account (see Fig.~\ref{fig:sv_mx}). 

Our bounds for XTE J1118+480 have a mild sensitivity on the effect of spatial diffusion (which implies at most a weakening of the bounds of a factor $\lesssim$ 6 at large $m_\chi$). On the other hand, diffusion is instrumental in enhancing the sensitivity of the results upon the intensity of the magnetic field (see Appendix~\ref{app:b_sens}).

In particular, we find that including the effect of diffusion a reduction factor of the magnetic field compared to the equipartition value larger than about 14 at low $m_\chi$, becoming progressively smaller at higher WIMP masses, would be sufficient to relax the XTE J1118+480 bound to the level of other existing bounds. Indeed, the recent study~\cite{dallilar_2018} of V404
Cygni, a black hole system quite similar to XTE J1118+480, leads to an estimation of the magnetic field about a factor of 20 smaller than its equipartition value. Although such determination is not conclusive, if it is representative of the possible excursion of the magnetic field close to a stellar--mass BH this would imply no WIMP bound at all from XTE J1118+480, with a predicted  signal below observation in the full $m_\chi$ range. We conclude that, although the bound from XTE J1118+480 on $\langle \sigma v\rangle$ is the most constraining compared to all others up to $m_\chi\simeq$ 1 TeV when the equipartition value for the magnetic field is assumed, the determination of the intensity of the magnetic field in BH–LMXB systems represents the major uncertainty in using them
to search for WIMPs as an alternative to heavier BHs.

\section*{Acknowledgements}
 The research of A K. , H, K. and S.S. was supported by the National Research Foundation of Korea(NRF) funded by the Ministry of Education
through the Center for Quantum Space Time (CQUeST) with grant number
2020R1A6A1A03047877 and by the Ministry of Science and ICT with grant
numbers 2021R1F1A1057119 and RS-2023-00241757. 

%\newpage
%\bibliographystyle{JHEP}
%\bibliography{bibliography}

\end{document}